\documentclass[%
groupedaddress,
amsmath,amssymb,
aps,
prx,
twocolumn,
notitlepage
]{revtex4-2}
\usepackage{graphicx}
\usepackage{dcolumn}
\usepackage{bm}
\usepackage[colorlinks=true,bookmarks=true,%
     pdfborder={0 0 0},linkcolor=blue,urlcolor=blue,%
     breaklinks=true,bookmarksnumbered=true]{hyperref}
\usepackage[stable]{footmisc}
\usepackage{wasysym}
\usepackage{times}
\usepackage{mathrsfs}
\usepackage{pdfsync}
\def \x{\bm{x}}

\def \s{\bm{s}}

\def \k{\bm{k}}

\def \i{\text{\bf{i}}}
\def \f{\text{\bf{f}}}

\def \hO{\hat{O}}
\def \ho{\hat{o}}
\def \ha{\hat{a}}
\def \hc{\hat{c}}

\def \hu{\hat{u}}
\def \hI{\hat{I}}
\def \hT{\hat{T}}
\def \hS{\hat{S}}
\def \hs{\hat{s}}
\def \heta{\hat{\eta}}
\def \hB{\hat{B}}
\def \hD{\hat{D}}

\def \tu{\tilde{u}}
\def \tT{\tilde{T}}

\def \tO{\tilde{O}}
\def \tG{\tilde{G}}

\def \si{\sigma}
\def \sib{\bar{\sigma}}

\def \Torder{\mathcal{T}}

\def \mH{\mathcal{H}}

\def \mT{\mathcal{T}}

\def \hilbert{\mathscr{H}}
\def \pt{\partial}

\def \tr{\text{Tr}}

\long\def\symbolfootnote[#1]#2{\begingroup               
\def\thefootnote{\fnsymbol{footnote}}\footnote[#1]{#2}   
\endgroup}                                               
\newcommand{\abs}[1]{\lvert#1\rvert}

\newcommand{\proj}[2]{\mbox{$|#1\rangle \!\langle #2 |$}}
\newcommand{\ev}[1]{\mbox{$\langle #1 \rangle$}}
\newcommand{\dev}[1]{\mbox{$\langle\langle #1 \rangle\rangle$}}
\newcommand{\cev}[1]{\mbox{$\langle #1 \rangle_c$}}

\newcommand{\ket}[1]{\mbox{$| #1 \rangle$}}

\newcommand{\ua}{\uparrow}
\newcommand{\da}{\downarrow}

\newcommand{\ancomm}[2]{\{ #1 , #2 \}}

\newcommand{\eqdisp}[1]{Eq.~(\ref{#1})}
\newcommand{\refdisp}[1]{Ref.~[\onlinecite{#1}]}

\begin{document}


\title{Algebraic-Dynamical Theory for Quantum Many-body Hamiltonians: \\
  A Formalized Approach To Strongly Interacting Systems
 }

\author{Wenxin Ding$^{1}$}\email{wxding@ahu.edu.cn}
 \affiliation{%
   $^1$School of Physics and Optoelectronics Engineering, Anhui University, Hefei, Anhui Province, 230601, China\\
 }%

\date{\today}

\begin{abstract}
Non-commutative algebras and entanglement are two of the most important hallmarks of many-body quantum systems. Dynamical perturbation methods are the most widely used approaches for quantum many-body systems. While study of entanglement-based numerical methods are booming recently, the traditional dynamical perturbation methods have not benefited from study of quantum entanglement.
In this work, we formulate an {\it algebraic-dynamical theory} (ADT) by combining the power of quantum algebras and dynamical methods in which quantum entanglement naturally emerges as the organizing principle. We start by introducing a {\it complete operator basis set} (COBS), with which an arbitrary state, either pure or mixed, can be represented by the expectation values of COBS.
Then we establish a complete mapping from a given state to a complete set of dynamical correlation functions of the state through the Heisenberg- and Schwinger-Dyson-equations-of-motion (SDEOM). The completeness of COBS and the mapping ensures ADT to be a mathematically complete framework in principle.
Applying ADT to many-body systems on lattices, we find that the quantum entanglement is represented by the cumulant structure of expectation values of the many-body COBS. The cumulant structure of the state forms a hierarchy in correlations. More importantly, such static correlational hierarchy is inherited by the dynamical correlations and their SDEOM. We propose that the dynamical hierarchy is also carried into any perturbative calculation on that state.
We demonstrate the validity of such perturbation hierarchy with an explicit example, in which we show that a single-particle-type perturbative calculation fails while a many-body perturbation following the hierarchy succeeds. We also discuss the computation and approximation schemes of ADT and its implications to other strong coupling theories like parton and slave particle methods.
\end{abstract}

\maketitle

\section{Introduction}
\label{sec:introduction}

Non-commutative algebra of quantum operators is one of the most important hallmarks of quantum mechanics, but for systems composed of more than one particle, quantum entanglement is an equally important characteristic. The importance of non-commutative algebra, as first proposed by W. Heisenberg in the matrix mechanics\cite{heisenberg-1925-matrix-mech}, was recognized at the birth of quantum mechanics. The first successful formalization of quantum mechanics was completed by von Neumann's via the $C*$-algebra\cite{book-von-Neumann-math-found-2018}. In contrast, the notion of quantum entanglement as ``a spooky action at distance'' was only pointed out in the seminal EPR paper\cite{EPR} ten years later.

The algebraic approach has long been used as a standard approach in textbooks of quantum mechanics to solve single-particle problems such as the quantum harmonic oscillator. W. Pauli\cite{Pauli1926}
derived the hydrogen atom spectrum algebraically in 1926 before the development of wave mechanics.
A comprehensive description of algebraic methods for quantum mechanics can be found in \refdisp{Adams1994}.

Algebraic methods for interacting few-body systems are developed in such as nuclear physics\cite{Casten1992} and molecule theory\cite{Iachello1995}, mostly along the line of Pauli's work: utilizing dynamic symmetry.
In extensive and interacting lattice systems, algebraic methods are also powerful, but mostly restricted to operator transformations. The bosonization method\cite{Gogolin1998}, the Jordan-Wigner transformation\cite{Jordan1928} and its generalizations\cite{Fradkin1989,Batista2001} are among the most widely-used algebraic transformations. 

Recently, quantum entanglement as an organizing principle for strong correlation in many-body systems was recognized with the development of the density matrix renormalization group (DMRG)\cite{White1992,scholl-dmrg-2005} method. Since then, more methods relying on entanglement, such as the matrix product state (MPS)\cite{verstraete-2007-mps}, tensor network states (TNSs) methods\cite{cirac-2009-renor-tensor}, etc., have been and are still being developed. While such developments greatly improved the understanding of many strongly interacting systems. But unfortunately, how entanglement can enter and improve the dynamical perturbation methods has not been investigated.

On the other hand, the dynamical perturbation theories\cite{AGD} have prevailed in calculating dynamical correlations and responses, both at zero and finite temperatures, for weak-coupling systems.
In strongly interacting systems, this approach runs into difficulties. The validity of the perturbation expansion relies crucially on Wick's theorem\cite{Wick1950}. For strongly interacting systems, i) the operators used to construct the Hamiltonian do not obey canonical commutation or anti-commutation relations (such as quantum spins), ii) or the unperturbed interacting limit is more conveniently described by noncanonical operators, such as the Hubbard operators for the Hubbard models\cite{Hubbard1967a,Hubbard1967,Hubbard1965,Hubbard1964,Hubbard1964a,Hubbard1963}. In both situations, the noncanonical operator algebras immediately invalidate Wick's theorem.
Even when working with canonical operators, the strongly correlations themselves also invalidate the Wick's theorem.
In some situations, it is possible to introduce a modified Wick's theorem as an approximation \cite{Tyablikov-1959,Vaks1968,Kondo1972} , which in some cases is also called ``random phase approximations'' (RPA). Such noncanonical theories were studied intensively for magnetically ordered systems\cite{Skryabin1988,Shimahara1991,Gasser2001,Frobrich2006,Majlis2014}. Recently, a general discussion of noncanonical degrees of freedom was given in \refdisp{quinn-2021-non-canon}.

Surprisingly, strong-coupling dynamical perturbation theories also face difficulties which are not well-understood. Taking the Hubbard model in the large onsite Coulomb repulsion (large-$U$) limit as an example. While a second order perturbation theory for two sites successfully accounted for the superexchange interaction among electronic spins in the half-filling limit, it also largely stops there. Attempts of perturbative calculations at finite doping run into various issues, such as negative spectral functions as found in \refdisp{pairault-2000-stron-coupl}.
Even in exactly solvable limits such as the dynamical mean field theory (DMFT)\cite{Georges1996} which is in the infinite spatial dimension limit, both numerical\cite{schaefer-2013-diver-precur,gunnarsson-2017-break-tradit,chalupa-2018-diver-irred} and analytic\cite{Thunstrom2018} calculations find singularities in the vertex functions, even for the simplest Hubbard atom problem. It is proposed that such singularities are potentially related to some of the difficulties that strong-coupling perturbation theories encounter.

Another common practice of treating strong correlations is to use parton or slave-particle constructions, all of which ``fractionalize'' the original physical operators into new operators of different properties. For example, quantum spins can be fractionalized into either fermions or bosons, such as like Abrikosov pseudo-fermions or Schwinger fermions\cite{abrikosov-1965-elect-scatt,affleck-1988-large-nlimit}, Schwinger bosons\cite{auerbach-2008-schwinger-boson}, Dyson-Maleev bosons\cite{Dyson1956,Maleev1958}, etc..
For Hubbard models, there are various versions of slave bosons\cite{kotliar-1986-new-funct,Florens2002}, slave spins\cite{DeMedici2005a}, slave fermions\cite{yoshioka-1989-slave-fermion} representations and the Hubbard operators\cite{Ovchinnikov-book-Hubbard-operator-2004}.
The choice of representation is often determined by problems of concern, since typically certain representations are more convenient in describing certain properties.

However, results of solvable models in one dimensional (1D) systems suggest that different types of excitations can coexist. For example, considering results found by the exact Bethe ansatz\cite{bethe-1931-zur-theor-metal} method in 1D quantum spin models, such as XXZ model in magnetic fields\cite{yang-2019-one-dimen}. When the spins are polarized by strong external magnetic fields, the system shows ferromagnetic (FM) spin-wave-like excitations, which is easily captured by bosonic partons. On the other end when magnetic field is turned off, a Luttinger-liquid-like spectrum appear, which is easier to capture for fermionic partons. In between, both spectral coexist. Besides, Bethe strings states\cite{takahashi-1999}, which is beyond the description of any type of partons, also come to play. While the string states typically live at higher energies, the gap of string excitations shrinks with the magnetization. Consequently, they contribute to the low energy spectrum when magnetization is small. In fact, they occupy more and more spectral weights towards the zero magnetization limit before suddenly disappearing when the field is turned off. For most of the parametric space with intermediate fields, all three types of spectral coexist and show complicated interplay behaviors.

While the boundary between fermions and bosons is considered obscure in 1D, such coexistence in higher dimensions is also present.
For example, in cuprates, there are a plethora of intertwined quantum orders of different natures\cite{Fradkin2012a,fradkin-2015-colloq}, either ``coexisting'' or ``competing''. Therefore, it is necessary to explore both the bosonic and fermionic algebras, which could shed light on theories in higher dimensions as the algebras are dimension-independent.

Recently, by combining the noncanonical Hubbard operator algebra and the dynamical perturbation method, B. S. Shastry developed the extremely correlated Fermi liquid (ECFL) theory\cite{Shastry2010, Shastry2011, Shastry2013, Shastry2013a} for the $t-J$ model\cite{Chao1977,zhang-1988-effec-hamil}. The ECFL theory solves for the electronic single particle Green's functions (GFs) with high accuracy at low energies, as benchmarked with DMFT\cite{zitko-2013-extrem-correl,perepelitsky-2016-trans-optic,ding-2017-stran-metal} and in one dimension with density matrix renormalization group (DMRG) calculations\cite{mai-2018-t-t} as well. Due to its analytic nature, ECFL is capable of studying two dimensional systems\cite{mai-2018-extrem-correl}.
The success of ECFL at solving for the low energy correlations with high accuracy shows the merits of combining the algebraic aspects with the dynamical approach.

In this work, by utilizing the full power of both the quantum algebras and the dynamical method, we formulate a {\it mathematically complete} framework which we call the {\it algebraic-dynamical theory} (ADT). ADT establishes a complete and consistent mapping between the states, either pure wave-functions or density matrices of mixed states, the many-body dynamical correlation functions via the Heisenberg-equations-of-motion (HEOM) and Schwinger-Dyson-equations-of-motion (SDEOM).
With such mathematically complete framework, we can further investigate controllable approximation schemes, such as a quantum statistical approach\cite{NNBogoliubov}. The analysis of application of ADT to a simple interacting system shows that the {\it quantum entanglement structures}, described by the many-body cumulant correlations, naturally emerges as the organizing principle. The cumulant correlations dictate a hierarchical structure among different levels of correlations\cite{aharonov-2018-compl-top}. Such static hierarchical structure is inherited by the dynamical correlations and any further calculations that are based them.


The rest of this work is organized as the following. In Sec. \ref{sec:algebr-dynam-appr}, we introduce the general framework of ADT and demonstrate the validity of ADT by solving a system in its spectral representation. In Sec. \ref{sec:algebr-dynam-theory}, we discuss the systems of interest to this manuscript, and their relations and mappings to a simple two-flavor problem in certain limits. Then we briefly discuss ADT solutions to this two-flavor problem, both exactly and perturbatively, to demonstrate that ADT can be used to do reliable perturbative calculations in the strongly interacting limit when the traditional single particle approach fails. After that, in Sec. \ref{sec:adt-lattice} we discuss ADT for lattice problems. We shall discuss implications of the two-flavor problem solutions to lattice problems. \ref{sec:adt-lattice} is devoted to the discussion of implications of ADT to problems encountered by other related theoretical methods. In the end, we summarize and discuss the future development and application of ADT.


\section{The Algebraic-Dynamical Approach}
\label{sec:algebr-dynam-appr}

In this section, we discuss the formulation of ADT, establishing the exact relations between the states and the dynamical correlation functions via HEOM and SDEOM for arbitrary Hamiltonian systems.


\subsection{The Complete Operator Basis Set}
\label{sec:compl-oper-basis}

In usual quantum mechanical problems, a Hamiltonian $\mH$ is given, alongside with a {\it complete set of commuting operators} (CSCO). The common eigenstate of the CSCO span the Hilbert space $\mathscr{H}$ of $\mH$. The physical observables are given as the expectation values of Hermitian operators in $\hilbert$.

In contrast, in order to have a complete description of the dynamics of a quantum system, it is necessary to extend the CSCO to a {\it complete  operator basis set} (COBS)\cite{Schwinger1960c}, which we denote as
\begin{align}
\mathscr{U} = \{ \hu^\alpha \}.\label{eq:1}
\end{align}
The elements of COBS generally do not commute, and can be non-Hermitian.
Instead, it is often convenient to also require $\mathscr{U}$ to satisfy the orthogonality condition\cite{Fano1957}
\begin{align}
  \label{eq:2}
  \tr ((\hu^\alpha)^\dagger \hu^\beta) = C ~\delta_{\alpha \beta},
\end{align}
where $C$ is a normalization factor which will be taken as 1 for convenience, unless noted otherwise. We also assume a Hermitian COBS unless note otherwise.
The completeness requires that the product of elements from COBS satisfy a set of {\it closed} algebraic relations
\begin{align}
  \hu^\alpha \hu^\beta = \sum_{\gamma} a^{\alpha \beta}_\gamma \hu^{\gamma},\label{eq:3}
\end{align}
where $a_{\alpha \beta}^\gamma \in \mathbb{C}$.
Satisfying both the complete and the orthogonal conditions, each operator $\hO$ of $\mathscr{H}$ can be expanded into a sum of $\hu^\alpha$s,
\begin{align}
  \label{eq:4}
  \hO = \sum_\alpha \tr (\hO \hu^\alpha) \hu^\alpha .
\end{align}

The closure of algebraic relations are crucial in uniquely specifying the underlying dynamical system. For example, for quantum spins-1/2, if only the $SU(2)$ Lie algebra is given, the total spin is not fixed. The usual spin-1/2 given represented by the Pauli matrices satisfying additional algebraic equations. Inversely, only when a complete set of algebraic relations is specified, the operator system is uniquely determined.

Conventionally, it is convenient to decompose the algebras into a symmetric (bosonic) sector and an anti-symmetric (fermionic) sector
\begin{align}
  [\hu^\alpha, \hu^\beta ] = \sum_{\gamma} b^{\alpha \beta}_\gamma \hu^\gamma, \quad
  \{ \hu^\alpha, \hu^\beta \} = \sum_{l} f^{\alpha \beta}_\gamma \hu^\gamma, \label{eq:5}
\end{align}
with
\begin{align}
   b^{\alpha \beta}_\gamma = a^{\alpha \beta}_\gamma - a^{\beta \alpha}_\gamma, \quad
  f^{\alpha \beta}_\gamma = a^{\alpha \beta}_\gamma + a^{\beta \alpha}_\gamma.\label{eq:6}
\end{align}


A few simple examples of COBS:
\begin{itemize}
\item for a single spin-1/2, one choice of orthogonal COBS is the set $\{\hI,  \si_x, \si_y, \si_z\}$;
\item for a harmonic oscillator, the COBS turns out to be an infinite set: $\{(\ha)^n (\ha^\dagger)^m,  (\ha)^m (\ha^\dagger)^n \}$ with $n,~m ~\in ~(\mathbb{N} ~\cup ~\{0\}) $.
\end{itemize}

Note that, in a many-body system, $\{ \hu_{\alpha} \}$ includes all many body operators and its size is exponentially large in system size $N$. For strongly interacting lattice systems, which are the focus of this work, the many-body operators can be constructed from the local $\hilbert_{i, \tau}$ of each site $i$ and flavor $\tau$. One straightforward construction would be taking the direct (tensor) product $\hT^{\alpha_i \alpha_j \dots}_{\tau_i \tau_j \dots ; ij\dots} = \hu^\alpha_{i \tau_i} \otimes \hu^\beta_{j \tau_j} \otimes \dots $, which are Cartesian tensor operators. However, such bases often are not convenient to use. More physical basis can be constructed as irreducible tensor operators from symmetry analysis, etc., which are widely know in angular momentum theories. Similar constructions should be employed to analyze systems on lattices.

\subsection{The States and the Density Matrix}
\label{sec:states}
Given a COBS, instead of specifying an arbitrary state as superposition of eigenstates or unit vectors of $\mathscr{H}$, the state can be specified by the expectation values of the bases operators $\ket{\{\ev{\hu^\alpha} \}}$ (as a vector). In particular, according to \eqdisp{eq:4}, the density matrix, pure or mixed, can be written as
\begin{align}
  \label{eq:7}
  \rho = \sum_\alpha \tr(\rho \hu^\alpha) \hu^\alpha = \sum_\alpha \ev{ \hu^\alpha} \hu^\alpha.
\end{align}
So the expectation value of any observable $\hO$ can expressed as
\begin{align}
  \label{eq:8}
  \ev{\hO} = \tr (\rho~\hO) = \sum_\alpha \ev{ \hu^\alpha} \tr (\hO \hu^\alpha).
\end{align}
Therefore, we argue that $\ket{\{ \ev{\hu^\alpha} \}}$ is a complete description for any state $\ket{\phi}$ as well as any density matrix $\rho$ which makes the description readily generalizable to finite temperatures.

Before going into the discussion of quantum dynamics, we note that the HEOM already puts nontrivial constriction on the state. Consider a Hamiltonian $\mH = \sum_{\alpha} h_\alpha \hu^\alpha$. Applying the HEOM to $\ev{\hu^\alpha}$, one obtains
\begin{align}
  \label{eq:9}
  \begin{split}
  & i \pt_t \ev{\hu^\alpha}  = \sum_\beta h_\beta \tr( \rho [\hu^\alpha, \hu^\beta] )= \sum_{\beta,\gamma}  b^{\alpha \beta}_\gamma h_\beta \ev{\hu^\gamma}.
\end{split}
\end{align}
At zero temperature, for an eigenstate, $\pt_t \ev{\hu^\alpha} = 0$ even if $[\hu^\alpha, H] \neq 0$ since any eigenstate does not evolve with time up to an overall phase factor. Therefore, \eqdisp{eq:9} becomes a constraining equation for $\ev{\hu^\alpha}$s. Systematically speaking, \eqdisp{eq:9} can be regarded as the SDEOM for the one-time-correlation-function $\ev{\hu^\alpha (t)}$, although no dynamics is involved for ground states or equilibrium states.
More interestingly, if the initial state is not an eigenstate, then \eqdisp{eq:9} would become dynamical.

\subsection{The Complete Set of (Two-time) Dynamical Correlation Functions}
\label{sec:CSDCF}

Given a COBS, we can consider the complete set of (two-time) dynamical correlation functions (CSDCF) $\mathbb{G}$. There can be different conventions of time-ordering in defining the elements $G$s. But in order to form a complete set, the equal time limit needs to cover the full algebra table of the COBS. In accordance to conventional time-order GFs for canonical bosons or fermions, we use the following conventions
\begin{align}
  \label{eq:10}
  i G_{\pm}[\hu^\alpha (t_i); \hu^\beta (t_f)]= \dev{\mT_\pm \Big( \hu^{\alpha} (t_i), \hu^{\beta} (t_f) \Big)},
\end{align}
where $\mT_\pm$ denotes time-ordering with the $\pm$ sign, $\dev{~}$ denotes fully dynamical correlations defined as
\begin{align}
  \label{eq:11}
    \dev{\hu^\alpha (t_i) \hu^\beta (t_f)} = \left\langle (\hu^{\alpha}(t_i)-\ev{\hu^\alpha}) (\hu^\beta(t_f) - \ev{\hu^\beta}) \right\rangle,
\end{align}
and $\hO (t) = U^\dagger(t) \hO U(t) $ is the time-evolved operator of the Heisenberg picture with $U(t) = e^{- i \mH t}$ being the usual time-evolution operator.
When taking the limit $t_i = t_f$, we have
\begin{align}
  i G_{+}[\hu^\alpha (t_i); \hu^\beta (t_i)] = \ev{ \{ \hu^{\alpha} , \hu^{\beta} \}} -2 \ev{\hu^{\alpha}} \ev{\hu^{\beta} }  ,\label{eq:12}\\
  i G_{-}[\hu^\alpha (t_i); \hu^\beta (t_i)] = \ev{ [ \hu^{\alpha},\hu^{\beta} ]}, \label{eq:13}
\end{align}
which are just taking the expectation values on all the algebraic relations in the form of \eqdisp{eq:5}.

$\bullet~$ {\it Simplify notations for GFs.}
From here on, we shall simplify our notation by writing all two-time correlation functions $ G_{\pm}[\hu^\alpha_i (t_i) \hu^\gamma_j (t_i) \dots ; \hu_l^\beta (t_f) \hu_m^{\eta}(t_f) \dots] $ as $ G_{\pm;ij\dots ; lm\dots} ^{\alpha\gamma\dots;\beta \eta \dots}[\i,\f] $, where the additional Latin letter indexes indicating spatial or momentum if any. When only two indexes are involved, the spatial indexes and the comma separating the operator indexes shall be ignored. Note that we use the bold-font $\i(\f)$ to denote the initial and final time. However, for the rest of this work, we always assume time-translational invariance, hence we also use $G_{\pm;ij\dots ; lm\dots} ^{\alpha\gamma\dots;\beta \eta \dots} [t]$ where $t = t_i - t_f$. Correspondingly, the frequency-Fourier transform of the GFs shall written as $G_{\pm;ij\dots ; lm\dots} ^{\alpha\gamma\dots;\beta \eta \dots}[\omega]$ since equilibrium condition is always assumed.


%

\subsection{The Schwinger-Dyson-equations-of-motion on CSDCF}
\label{sec:SDEOM-CSDCF}
When given a COBS $\mathscr{U}$ and a state or a density matrix, specified in the form of $\ket{\ev{\mathscr{U}}}$, we can solve for the CSDCF by taking the time-derivative on all the elements of CSDCF and apply the HEOM:
\begin{align}
  \label{eq:14}
  \begin{split}
    i \pt_{t_i } G_{\pm}^{\alpha ;\beta}[\i,\f] & = \delta^{\alpha \beta}_{\pm} [\i,\f] + i \dev{\mT_\pm \big( [\hu^\alpha(t_i) , \mH], \hu^\beta(t_f) \big) }\\
    & = \delta^{\alpha \beta}_{\pm}[\i,\f] +  \sum_{\eta \gamma} h_\eta  b^{\alpha \eta}_\gamma G_{\pm}^{\gamma ;\beta}[\i,\f],
\end{split}
\end{align}
where
\begin{align}
  \delta^{\alpha \beta}_{\pm}[\i,\f] = \delta(t_i - t_f)  i G^{\alpha; \beta}_{\mp}[\i,\i]
  ,\label{eq:15}
\end{align}
The time-derivative of the dynamical $G^{\alpha ; \beta}_{+(-)} [\i,\f]$s are dependent of the equal-time correlators $G^{\alpha ; \beta}_{-(+)}[\i,\i]$. This is in agreement with our assertion that $G^{\alpha ; \beta}_{\pm} [\i,\f]$s are independent in general, but with exceptions. When time-translation-invariance is present, the equations can be cast to the frequency space and be solved as a set of linear equations.

Therefore, we can arrange the CSDCF into a vector form, and rewrite the complete set of SDEOM of the CSDCF in a matrix form:
\begin{align}
  \label{eq:16}
   i \pt_{t_i } [{\mathbf G}_{\pm}] = [\mathbf{\Delta}]_{\pm} [\i,\f] + [[\mathbf{L}]] \cdot [{\mathbf G}_{\pm}],
\end{align}
where we use $[\mathbf{\Delta}]_{\pm}$ as the vector form of $\{ \delta^{\alpha \beta}_{\pm} [\i,\f] \}$ (with ${\alpha}$ as the component index) and $ [[\mathbf{L}]] $ as the matrix form of $\{ \sum_{\eta} h_\eta  b^{\alpha \eta}_\gamma \}_{\alpha \gamma}$ (with different $ \alpha$ and $\gamma$ as the element indexes).

The resulting \eqdisp{eq:14} are the many-body version of the SDEOM. In Schwinger's formulation for interacting theories, for the terms  on the right-hand-side (RHS) other than the original $G^{\alpha;\beta}$, interactions often introduce additional operators, turning $G^{\alpha;\beta}$ into vertex functions of the form $G^{\alpha \gamma; \beta}$. Schwinger introduced the source field technique which turns such vertex functions into functional derivatives with respect to the added operators such as
\begin{align}
  \label{eq:17}
  G_\pm[\hu^\alpha (t_i) \hu^\gamma(t_0); \hu^\beta (t_f)] = \frac{\delta}{\delta \mathcal{V}_{\hu^\gamma}(t_0)} G_\pm^{\alpha; \beta}[\i,\f],
\end{align}
in which a source term\cite{schwinger-source-tech} $\mathcal{A} = \sum_{\gamma} \int dt \mathcal{V}_{\hat{u}_\gamma}(t) \hat{u}_\gamma$ must be added. All sources are set to zero in the end. The vertex functions are more generic as the inserted operators are allowed at an arbitrary time. In this work, we restrict to the two-time cases, which limits $t_0 \rightarrow t_i$. Despite the restriction, the completeness of COBS and CSDCF formally ensures the solvability of ADT.

\subsection{Algebraic-Dynamical Theory in the Spectral Representation}

For a given $\hilbert$, one immediate COBS can be constructed from its spectral representation.
Denoting the eigenvalues and eigenstates of $\mH$ as $\{\epsilon_n,~\ket{n,~g_n} \}$ with $n=1,2,\dots,N$, where $g_n$ denotes any possible degeneracy-index.
For simplicity, we consider the case where no degeneracy is present, i.e. $\forall~n$, $g_n = 1$, and ignore the $g_n$ index.
However, the presence of degeneracy does have very important consequences which will be discussed later.

$\bullet~$ {\it The COBS is given by $\{ \hO_{mn} \}$}, where $\hO_{mn} = \proj{m}{n}$. Obviously, an arbitrary operator within the Hilbert space of $\mH$ can be expressed as superposition of $\hO_{mn}$s.

$\bullet~$ {\it The Hamiltonian:} $\mH = \sum_n \epsilon_n \hO_{nn}.$

$\bullet~$ {\it The algebras of the spectral operator basis:}
\begin{eqnarray*}
  [\hO_{mn}, \hO_{lk}]_{\pm} = \delta_{nl} \hO_{mk} \pm \delta_{mk} \hO_{ln}.
\end{eqnarray*}

$\bullet~$ {\it The commutation relation with $\mH$, i.e. the HEOM:}
\begin{eqnarray*}
  [\mH, \hO_{mn}] = (\epsilon_m - \epsilon_n) \hO_{mn}.
\end{eqnarray*}

$\bullet~$ A state $\ket{\Psi} = \sum_n c_n \ket{n}$ is given by the expectation values of $\hO_{mn}$'s
\begin{eqnarray*}
  \ev{ [\hO_{mn}, \hO_{nm}]_{\pm} } = \abs{c_n}^2 \pm  \abs{c_m}^2,\\
  \ev{ [\hO_{ml}, \hO_{ln}]_{\pm} } = c_m^* c_n, \ev{ [\hO_{lm}, \hO_{nl}]_{\pm} } = c_m^* c_n.
\end{eqnarray*}

%

$\bullet~$ {\it The complete set of Schwinger-Dyson-Equations-of-Motion Hierarchy:}
  \begin{align}\label{eq:18}
    \begin{split}
      & i \pt_{t_i}  G _{\pm}[\hO_i(t_i), \hO_f(t_f)] = \delta(t_i - t_f) \dev{[\hO_i, \hO_f]_{\mp}} \\
      &  + (\epsilon_m - \epsilon_n) G _{\pm} [\hO_i(t_i), \hO_f(t_f)],
    \end{split}
  \end{align}
where we take $\hO_i = \hO_{mn}$ and the HEOM have been applied. Both $\hO_i$ and $\hO_f$ should run through the full COBS. The SDEOM is automatically closed in spectral representation thus immediately solved.

$\bullet~$ {\it Solution in frequency space:}
\begin{align}
 G_{\pm} [\hO_{\i}, \hO_{\f}] [\omega] = \frac{\dev{[\hO_{\i}, \hO_{\f}]_{\pm}}}{\omega - (\epsilon_m - \epsilon_n)}.\label{eq:19}
\end{align}
Given all solutions of $G[\hO_{\i}, \hO_{\f}]$ for $\hO_i$ \& $\hO_f \in$ COBS, any other $G$s can be computed as their superposition.


\section{Algebraic-Dynamical Theory for Simple Limits}
\label{sec:algebr-dynam-theory}
For a large class of strongly interacting systems, the model Hamiltonians are defined on a lattice with a finite flavor index (like spin, orbit, etc.). For every lattice point $\x_{i}$ and flavor there is a local Hilbert space $\mathscr{H}_{\tau, i}$, where $\tau$ tracks all flavors. We can denote the local and single flavor COBS as $\mathscr{U}_{\tau, i} = \{ \hu^{\alpha}_{\tau, i} \}$.

Just as the many-body Hilbert space $\mathscr{H}$ can be constructed as direct product $\otimes_{\tau,i} \mathscr{H}_{\tau, i}$, the many-body COBS can also be constructed out of the local COBS $\otimes \mathscr{U}_{\tau, i}$. The many-body COBS can be classified by how many different flavor-site they involve, which is essentially their ``grades''. For a given grade-$n$ with a fixed set of $n$ flavors and/or sites $\{(\tau, i), (\tau', j),\dots\}$, the grade-$n$ COBS can be constructed by Cartesian products. For example, $n = 1, 2, 3$ are given below
\begin{eqnarray}
  \label{eq:20}
  \begin{split}
    \text{grade 1:} \{ \hu^{\alpha}_{\tau, i} \},\\
    \text{grade 2:} \{\hT^{\alpha \beta}_{2:\tau\tau';ij} = \hu^{\alpha}_{\tau, i} \hu^{\beta}_{\tau', j}\},\\
    \text{grade 3:} \{ \hT^{\alpha \beta \gamma}_{3:\tau\tau'\tau'';ijk} = \hu^{\alpha}_{\tau, i} \hu^{\beta}_{\tau', j}  \hu^{\gamma}_{\tau'', k}\},\\
    \dots \dots\dots\dots .
  \end{split}
\end{eqnarray}
However, there are some important subtleties in the above definition.
\begin{itemize}
\item Since grade-1 COBS $\mathscr{U}_{\tau,i}$ contains the identity operator $\hI_{\tau,i}$, the grade-2 COBS also contains the grade-1 operators as $\hI_{\tau,i} \hu^\alpha_{\tau',j}$, which we shall the ``single particle(spin) operators'' in a many-body systems.
\item Algebraically, for grade-$n$ with $n>1$, it is often more convenient to use linear combinations similar as the irreducible tensor operators. For example, for two quantum spins, we would prefer to use the total spin operator $\hS^\alpha_{ab} = \hS^\alpha_a + \hS^\alpha_b$ and a staggered spin operator $\hat{\eta}^\alpha_{ab} = \hS^\alpha_a - \hS^\alpha_b$ instead of the single spin operators $\hS^\alpha_{a(b)}$s when studying the correlation functions at grade-2.
\item Therefore, instead of the simple Cartesian product construction, we always use a more generic construction which makes linear combinations of the  Cartesian product operators. For example, for grade-2 operators, we introduce
  \begin{align}
    \{\hT^{\gamma}_{2:\tau\tau' , ij} = \sum_{\alpha \beta} t^\gamma_{\alpha \beta} \hu^{\alpha}_{\tau, i} \hu^{\beta}_{\tau', j}\}\label{eq:21}
  \end{align}
\end{itemize}
instead of \eqdisp{eq:19}.
This is similar to the construction of spherical tensor operators in normal quantum mechanics.

The many-body Hamiltonians we are interested in are local or short ranged, including both  interactions and hopping. Therefore, we prefer to organize a Hamiltonian by locality and the grades:
\begin{eqnarray}
  \label{eq:22}
  \mH = \sum_i (\mH_{1,i} + \mH_{2,i})+ \sum_{i \neq j} \mH_{2,ij}  + \dots,
\end{eqnarray}
where
\begin{eqnarray*}
  \mH_{1,i} = \sum_{\alpha, \tau} h^\alpha_{\tau} \hu^\alpha_{\tau,i},\\
  \mH_{2,i} = \sum_{\alpha\beta, \tau \neq \tau'} J^{\alpha\beta}_{\tau\tau',ii} \hT^{\alpha \beta}_{2: \tau \tau',i i},\\
  \mH_{2,ij} = \sum_{\alpha\beta, \tau\tau'} J^{\alpha\beta}_{\tau\tau' , ij} ,\hT^{\alpha \beta}_{2: \tau\tau', ij},
\end{eqnarray*}
We restrain the scope of this work to up to grade-2 operators although higher order interactions could be important in many cases and real systems.

In \eqdisp{eq:19}, these are composite tensor operators on the lattices . One of the most widely used ones are the spin operators $\hs^\alpha_i = \sum_{\si \si'} c^\dagger_{\si} \si^\alpha_{\si \si'} c_{i \si'} $ for electrons in a Mott insulating state. We must note that, the grade of operators depends on the underlying system of concern. For example, in pure quantum spin systems, $S^\alpha_i$s are grade-1, while in electronic systems they become grade-2.
We should emphasize that, the other composite tensor operators should be considered no less physical if defined properly.

\subsection{Models of Interest}
\label{sec:models-interest}

In this work, we focus three classes of strongly interacting models, namely i) the quantum spin-1/2 models (QSMs), ii) Hubbard models(HMs) and iii) Kondo type models, including impurity or lattice Kondo models(KMs).
For simplicity, we restrict our discussion to single orbital cases (except for KMs), therefore, orbital index $\tau$ shall be suppressed.

All these three classes of models are composed of the quantum spin-1/2 operators, $\{ \hs^\alpha_i \}$(QSMs), or the canonical fermionic creation/annihilation operators $\{c_{i \si}, c^\dagger_{i \si}\}$ (HMs), or both (KMs).
However, the composition units, i.e. the COBS of a single site and flavor, satisfies a  $SU(2)$ Lie algebra and a Clifford algebra $Cl(2,0)$ at the same time. Equivalently, the complete algebraic relations would dictate that the underlying operators must be spin-1/2.

For spin-1/2, we have
\begin{eqnarray}
  [\hs^{\alpha}_i, \hs^{\beta}_i] = i \epsilon_{\alpha \beta \gamma} \hs^{\gamma}, \label{eq:23}\\
  \{ \hs^+_i, \hs^-_i\} = 1, \quad \ancomm{\hs^\pm}{\hs_i^z} = 0.\label{eq:24}
\end{eqnarray}
The local Hamiltonians of the most interest are perhaps the Heisenberg-type interaction $\mH_{2;ij} = \sum_{\alpha} J^\alpha_{ij} \hs^\alpha_i \hs^\alpha_j$.

For canonical fermions of a single flavor, letting
\begin{eqnarray}
  \begin{split}
    \gamma^{x}_{i\si} = \frac{ c_{i \si}  + c^\dagger_{i \si} }{2},
    \gamma^{y}_{i\si} = \frac{ c_{i \si}  - c^\dagger_{i \si} }{2i}, \\
    \gamma^z_{i \si}= c^\dagger_{i \si} c_{i \si} - \frac{1}{2},
  \end{split}
  \label{eq:25}
\end{eqnarray}
we have
\begin{eqnarray}
  [\gamma^{\alpha}_{i\si}, \gamma^{\beta}_{i\si}] = i \epsilon_{\alpha \beta \gamma} \gamma^{\gamma}_{i\si}, \label{eq:26}\\
  \{ \hc_{i \si}, \hc^\dagger_{i \si}\} = 1, \quad \ancomm{\hc^\dagger}{\gamma^z} = 0, \quad \ancomm{\hc}{\gamma^z} = 0.\label{eq:27}
\end{eqnarray}
The $\gamma^\alpha_{i \si}$'s are the Majorana fermion operators for canonical fermions.

Therefore, for all three types of models, there is a unified ``local'' limit involving two parties. In terms of the single-particle COBS, we have the following.
\begin{itemize}
\item For single-flavor QSMs, we consider the interactions \& fields on a single bond, i.e. a pair of sites $(a,b)$;
  \begin{align}
    \label{eq:28}
    H_{QSM,ij} = \sum_{\alpha,l} h^\alpha_{l} \hs^\alpha_l + \sum_{\alpha \beta} J^{\alpha \beta} \hs^\alpha_a \hs^\beta_b.
  \end{align}
\item For single-band HMs, that is just the electron on a single site with the opposite spins, also known as the Hubbard atoms
  \begin{align}
    \label{eq:29}
    H_{HM, i} = \sum_\si \mu_\si^\alpha  \gamma^\alpha_\si + \frac{U}{2} \gamma^z_\si \gamma^z_{\sib};
  \end{align}
  Here $\mu^z_{\si}$ is the chemical potential, while $\mu^{x(y)}_{\si}$ can be interpreted as a mean-field term stems from hopping/pairing.

\item For KMs, that is a minimal model with  one electron interacting with a single local moment:
  \begin{align}
    \label{eq:30}
    H_{KM, 0} = \sum_\si \mu_\si^\alpha  \gamma^\alpha_\si + \sum_{\alpha} h^{\alpha} \hs^\alpha + \sum_{\alpha} J_K \hs^\alpha \hS_c^\alpha,
  \end{align}
  where $\hS_c^\alpha = \sum_{\si \si'} c^\dagger_{\si} \si^\alpha_{\si \si'} c_{\si'}$.
  For KMs, the overall Hilbert space is larger, however, the interaction between the fermions is excluded, the nontrivial dynamics is between the electronic spin and the local moment. Thus the discussion is equally applicable.

\item The Anderson impurity models and periodic Anderson models are multi-orbital generalizations of the Hubbard models, thus are also within the scope of this paper.

\item Descendants of the above models in the strong coupling limit, such $t-J$ models, where only a subset of operators are kept. The Kondo models can be considered as descendants of the Anderson impurity model or the periodic Anderson models.

\end{itemize}

Therefore, we focus our discussion on a series of generic QSM Hamiltonians with such local algebraic equivalent relations. 
For example, the Hubbard-$U$ interaction $\frac{U}{2} \gamma^z_{\si} \gamma^z_{\sib}$ can be considered equivalent to an Ising interaction term $J^z \hS^z_i \hS^z_j$, $\gamma^x_\si \gamma^x_{\sib}$ can be considered equivalent to $(\hc^\dagger_\si + \hc_\si) (\hc^\dagger_\si + \hc_{\sib})$ and etc..

Such local algebraic-equivalence is because these systems all have a $2$-dim Hilbert space for a single flavor. When lattice index is included, the algebraic equivalence is no longer valid. However, it is still possible to construct operator-transformations which essentially transmutes the spatial statistics while preserving the local algebraic structure\cite{Batista-Ortiz-2004}. In one-dimension, such transmutation of statistics leads to methods like Jordan-Wigner transformation (JWT), bosonization, etc.. In higher dimensions, similar constructions generally have much less power.

Next, we discuss the ADT solution of the local limits within this work as a proof of principle. Detail studies involving the spatial terms will be presented separately in the future.


\subsection{Solutions of the Two-flavor Problem}
\label{sec:solut-local-limits}

In this part, we briefly describe the exact solutions of the local limits discussed in Sec.~\ref{sec:models-interest} to shed light on our final goal: solving the lattice problems. We use the QSMs for the discussion and refers to algebraic-equivalence in Sec.~\ref{sec:models-interest} for other models of interests.

Consider an QSM Hamiltonian
\begin{align}
  \label{eq:31}
  \mH_{ab} = \mH_{0,ab} + \mH_{1,ab},
\end{align}
where $  \mH_{0,ab} = J^z \hs^z_a \hs^z_b + h^z (\hs^z_a + \hs^z_b)$, $\mH_{1,ab} = h^x (\hs^x_a + \hs^x_b)$.
The Ising interaction term $J^z \hs^z_a \hs^z_b$ is considered as the ``free theory'', and since the $h^z$ term commutes with the Ising term. A transverse field term $\mH_1$ is taken as a perturbation. If we consider $J^z<0$, this toy model can be extended to a lattice as the transverse field Ising model when $h^z = 0$. When $J^z>0$ this model can be viewed as the atomic limit of the HM, as previously discussed while the $\mH_1$ can be interpreted as a Weiss mean field term stemming from the hopping. For the rest of this section, we only consider $J^z >0$ without losing generality.

Even for such simple limits, the many-body dynamical correlations are recently found to show singularity in the context of a Hubbard atom \cite{Thunstrom2018} or a Kondo impurity\cite{Chalupa2021}. As we shall show later, consistent results are found at zero temperature.

The ground state $\ket{\Psi_0}$ can be characterized as the following.
\begin{itemize}
\item When $h^x$ is kept $0$,
  \begin{itemize}
  \item $\abs{J^z} < \abs{h^z}$, $\ket{\Psi_0} = \ket{\uparrow \uparrow}$ or $\ket{\downarrow \downarrow}$, fully polarized and unentangled, depending on the sign of $h^z$;
  \item $ J^z > \abs{h^z}$, $\ket{\Psi_0} = c_1 \ket{\uparrow \downarrow} + c_2 \ket{\downarrow \uparrow}$, a ground state can be an arbitrary vector inside the 2-dimensional degenerate subspace;
  \item  level-crossings happens across the ``quantum critical points'' where $\abs{J^z}  = \abs{h^z}$.
  \end{itemize}
\item When $h^x \neq 0$, the above three regimes still exist approximately, but smooth crossovers happen in between; the transverse field $h^x$ can always induce a transverse magnetization as long as it is nonzero. It is those crossover regimes that are of particular interest to us.
\item We are mostly interested in distinction between the product states and the entangled states. We shall focus on $ J^z > \abs{h^z}$, where without $h^x$, the ground state(s) form a 2-dimensional subspace and turning on $h^x \rightarrow 0^+$ immediately removes the degeneracy making the triplet Bell state $ (\ket{\ua \da} + \ket{\da \ua})/\sqrt{2} $ the ground state.
\item We shall focus on $h^x = 0$, making a distinction between the product state and the Bell state. Then we use the $h^x=0$ results as the ``free theories'' to perturbatively deduce the ground state for $h^x \rightarrow 0^+$.
\end{itemize}

\subsubsection{The grade-2 description of ``free theories''.}
\label{sec:grade-2-cobs}

To provide a ADT solutions, we first specify a grade-2 COBS for the problem. Since there are only two site involved, grade-2 is also the largest possible grade for the problem.
Consider two spin-1/2's $\s_a$ \& $\s_b$.  We use the following grade-2 COBS construction:
\begin{align}
  \begin{split}
    &\hS^\alpha_{ab} = \hs_a^\alpha + \hs_b^\alpha,~
    \heta_{ab}^\alpha = \hs_a^\alpha - \hs_b^\alpha,\\
    & B_{A,ab}^{\gamma} =2  \sum_{\alpha,\beta}\varepsilon_{\alpha \beta \gamma} \hs_a^\alpha \hs_b^\beta ,\\
    & B_{S,ab}^{\gamma} =2  \sum_{\alpha,\beta} \varepsilon_{\alpha \beta \gamma}^2 (\hs_a^\alpha \hs_b^\beta + \hs_a^\beta \hs_b^\alpha),\\
   &  D_{ab}^\alpha = 2 \hs_a^\alpha \hs_b^\alpha,
  \end{split}
      \label{eq:32}
\end{align}
Although a direct product construction is equally valid and more convenient to implement in programming, this construction provides more physical insight which we shall discuss later. This particular construction follows from the geometric algebra\cite{Doran2009}. But for the discussion of dynamical correlation functions, we stick to the Cartesian COBS.
Since $B^\gamma_{S(A)ba} = \pm B^\gamma_{S(A)ab}$, we drop the $_{ab}$ index and denote $B_{S(A)}^\gamma = B^\gamma_{S(A)ab}$. So the grade-2 COBS is written as
\begin{align}
  \label{eq:33}
  \mathscr{U}_{g2} = \{\hI, \hS^\alpha, \heta^\alpha, \hB_S^\alpha, \hB_A^\alpha, \hD^\alpha \}.
\end{align}
Any state should be described by $\ev{\mathscr{U}_{g2}}$, which has 16 real numbers, exceeding the 8 real parameters allowed by considering the 4-dimensional $\hilbert_{ab}$. This reflects the algebraic constraints on the operators' expectation values. If the state is an eigenstate, additional constraints arises through the HEOM. For simplicity, we only discuss two extreme cases: i) the product states and ii) the Bell states $\ket{t/s} = \frac{1}{\sqrt{2}} (\ket{\ua \da} \pm \ket{\da \ua})$. In order to better characterize the many-body correlations, we also consider the cumulant expectation values $\dev{~} $ for composite operators. For grade-2 operators, it is defined as
\begin{eqnarray}
  \ev{\hT^{\alpha \beta}_{2:\tau\tau' , ij}} = \dev{\hT^{\alpha \beta}_{2:\tau\tau' , ij}} + \ev{\hu^{\alpha}_{\tau, i}} \ev{\hu^{\beta}_{\tau', j}},\label{eq:34}
\end{eqnarray}
where
\begin{eqnarray}
  \label{eq:35}
  \begin{split}
    \dev{\hT^{\alpha \beta}_{2:\tau\tau' , ij}} = \ev{\hT^{\alpha \beta}_{2:\tau\tau' , ij}} - \ev{\hu^{\alpha}_{\tau, i}} \ev{\hu^{\beta}_{\tau', j}} \\
    = \ev{(\hu^{\alpha}_{\tau, i}  - \ev{\hu^{\alpha}_{\tau, i}})(\hu^{\beta}_{\tau', j} - \ev{\hu^{\beta}_{\tau', j}})}.
  \end{split}
\end{eqnarray}
Now we examine $\dev{\mathscr{U}_{g2}}$ for these two cases.
\begin{itemize}
\item For a AFM product state $\ket{\ua \da}$ or $\ket{\da \ua}$, only a single particle operator acquire non-zero expectation value $\dev{\heta^z} = \pm 1$.
\item For the entangled state $ (\ket{\ua \da} \pm \ket{\da \ua})/\sqrt{2}$, the only nonzero ones are $\dev{\hD^x} = \pm 1/2,~\dev{\hD^y} = \mp 1/2,~\dev{\hD^z} = -1/2$.
\end{itemize}

Now following our ADT prescription, we can solve for all the  functions. However, even for such simple limit, the overall number of correlation functions is over 200, but with a large redundancy. While the full and exact solutions to this problem will be presented separately\cite{wxding-2022b}, here we want to begin with the single particle GFs. In a conventional sense, the single particle GFs are $G^{+;-}_{+;i;f}[\i,\f]$ and $G^{-;+}_{+;i;f}[\i,\f]$ for spins. 
For example, taking the time-derivative on  and applying the HEOM leads to
\begin{equation}
  \label{eq:36}
  \begin{split}
      i \pt_{t_i} G^{+-}_{+}[\i,\f] &= \delta(t_i-t_f) \delta_{i,f}\ev{2 \hs^z_i} \\
    & - J^z G^{+z;-}_{+;ij;f}[\i,\f] + h^z G^{+-}_{+}[\i,\f].
  \end{split}
\end{equation}

For strong interactions, the key is to understand the behavior of $i G^{+z;-}_{+;ij;f}[\i,\f] = \dev{\Torder_+ \left[ \hs^+_{i}(t_i) \hs^z_{j} (t_i), \hs_{f}(t_f) \right]}$.
For example, our exact solutions for $G^{+;-}_{+;a;a}[\omega]$ and $G^{+z;-}_{+;ab;a}[\omega]$ with a $\ket{\Psi_0} = \sin\theta \ket{\ua \da} + \cos\theta \ket{\da \ua}$ read
\begin{align}
  \label{eq:37}
  G^{+-}_{+;aa}[\omega] = \frac{ J^z \ev{\hs^z_a} }{(\omega + J^z/2) (\omega - J^z/2)}, \\
  \label{eq:38}
  G^{+z;-}_{+;ab;a}[\omega] = \frac{\omega \ev{\hs^z_a} }{(\omega + J^z/2) (\omega - J^z/2)}.
\end{align}
\subsubsection{Transverse field as perturbation}
\label{sec:transv-field-pert}
On top of the above ``free theory'', the ADT approach allows for perturbative calculations by utilizing the complete set of algebraic relations. For this two-site problem, an important perturbation would be turning on the transverse field. Starting from the solutions for $h^x =0$, we make use of the following relations
\begin{align}
  \label{eq:39}
  & \ev{\hs^x_a} = -i \ev{[\hs^z_a, \hs^y_a]}  =  G_{-;aa}^{zy}[\i,\i].
\end{align}
Without $h^x$, $G^{(0)zy}_{-;aa} = 0$.
When $h^x$ is turned on, a new contribution $\propto h^x$ to HEOM $i \pt_t \hs^z_i  = i h^x \hs^y_i$. Thus a naive way to do the perturbation is to follow the corresponding new contribution to SDEOM as
$\omega G^{zy}_{-;aa}[\omega]  =  i h^x G^{yy}_{-;aa}[\omega] \simeq i h^x G^{(0);yy}_{-;aa}[\omega]$,
where in the second equation a perturbative iteration to the first order in $h^x$ is used. We find $$\ev{\hs^x_a} = \frac{ 4 h^x h^z }{(J^z)^2 - 4 (h^z)^2}  \ev{\hs^z_a}_0,$$ which agrees with diagonalization study well at sizable $h^z$ and $\ev{\hs^z}$ at the leading order of $h^x$. However, this perturbative calculation fails as $h^z \rightarrow 0$ and $\ev{\hs^z} \rightarrow 0$. In contrast, the exact result is $ \ev{\hs^x_i} \simeq - h^x/J^z$ even when $h^z =0$ \& $\ev{\hs^z} = 0$. In fact, the susceptibility $\chi_{h^x} = \pt_{h^x}\ev{\hs^x_i}$ is 2 times of that when only $h^z$ is applied. The presence of strong fluctuations in fact enhances the susceptibility.

In fact, after a careful scrutiny of the full set of SDEOM, we find it impossible to obtain a proper perturbation starting from the singlet state with the relation \eqdisp{eq:39}. The SDEOM keeps loop within a trivial set of correlation functions. The underlying reason is that the single particle operators form a subgroup and cannot generate elements beyond its subgroup.


To achieve a proper perturbation theory, we need to start with a many-body analogue of \eqdisp{eq:39}
\begin{align}
  \label{eq:40}
  \begin{split}
    [\hs^z_a \hs^z_b, \hs^y_a \hs^z_b] = i \hs^x_a (\hs^z_b)^2
    \rightarrow \ev{\hs^x_a} = - 4 i \ev{[\hs^z_a \hs^z_b, \hs^y_a \hs^z_b]},
  \end{split}
\end{align}
where $(\hs^z)^2 = 1/4$ is applied.
Now we can utilize the following SDEOM (higher order terms in $h^x$ are ignored)
\begin{align}
  \label{eq:41}
  i \pt_{t_i} G ^{zz;yz}_{-;ab;ab}[\i;\f] \simeq  - i h^x G ^{zy;yz}_{-;ab;ab} [\i;\f]
\end{align}
to immediately give a proper perturbation calculation for $\ev{\hs^x_a}$
\begin{align}
  \label{eq:42}
  \begin{split}
    \ev{\hs^x_a} &=  - 4 h^x G_- ^{zz;yz}[\i;\i] \\
    & = -4  h^x \int \frac{d\omega}{\pi} ~ i \frac{G_{-;ab;ab} ^{(0)zy;yz}[\omega]}{\omega} = \frac{- 4 h^x \ev{\hs^y_a \hs^y_b} }{J^z}.
  \end{split}
\end{align}
For the singlet, $\ev{\hs^y_a \hs^y_b} = 1/4$. The exact numerical value $-2 h^x/J^z$. The extra factor of 2 is due to a correction to the vertex function on RHS of \eqdisp{eq:41}.

\subsubsection{Analysis from ADT perspective}
\label{sec:analysis-from-adt}

To understand the calculation better, we analyze the results of the previous section with ADT formalism.

For the unperturbed, two-fold-degenerate ground state, we can parameterize it as $\ket{\Psi_0} = \sin\theta \ket{\uparrow \downarrow} + \cos\theta \ket{\downarrow \uparrow}$. Such a state can be represented by expectation values of COBS as shown in Table \ref{tbl:state}.
\begin{table}[htbp]
\centering
\begin{tabular}{|c|c|}
\hline
$\ev{\hs^z_{a(b)}}_0$ & $\pm \cos(2 \theta)/2$\\
\hline
$\ev{\hs^x_a \hs^x_b}_0 = \ev{\hs^y_a \hs^y_b}_0$ & $\sin(2\theta)/4$\\
\hline
$\ev{\hs^z_a \hs^z_b}_0$ & $-\sin^2(2\theta)/4$\\
\hline
\end{tabular}
\caption{COBS representation of the ground state. Zero elements are ignored.}
\label{tbl:state}
\end{table}

In the frequency space, we can write Eq. \eqref{eq:16} as
\begin{equation}
  \left( \omega - [\mathbf{L}] \right) \cdot [{\mathbf G}_{\pm}] = [\mathbf{\Delta}]_{\pm},
  \label{eq:43}
\end{equation}
which is a system of linear equations of $G$s in a matrix form. To solve it, one has to pivot the equations first. Apparently, the important pivot elements here are $[\mathbf{\Delta}]_{\pm}$. For the triplet state $\theta=\pi/4$, we identify the relevant $G$s through algebras associated with the only nonzero elements $\ev{\hs^\alpha_a \hs^\alpha_b}_0$
\begin{align}
  \label{eq:44}
   [\hs^x_a, \hs^y_a \hs^z_b] = 2 i \hs^z_a \hs^z_b,~[\hs^x_a, \hs^z_a \hs^y_b] = - 2 i \hs^y_a \hs^y_b \dots,
\end{align}
and the corresponding $G$s and their SDEOM (by taking the time-dependent averages on the algebraic equations). They are the {\it pivotal equations} of Eq. \eqref{eq:43}. When we solve a linear system, we begin with solving the pivotal equations first. In other words, solutions of non-pivotal $G$s, i.e. with zero $\mathbf{\Delta}_\pm$, originate from these pivotal $G$s. We shall call such pivotal $G$s the {\it parental channels}. Similarly, perturbations should also begin with the pivotal equations.

Therefore, we can understand the difference between Eq. \eqref{eq:39} and \eqref{eq:42} as the following. With the triplet state, $\delta\ev{\hs^x_a} \leftrightarrow h^x G^{(0) y;y}_{-;a;a} \leftrightarrow G^{(0);x;yz}_{-;a;ab} \propto \ev{\hs^z_a \hs^z_b}_0$. Therefore, Eq. \eqref{eq:39} can be understood as a contribution of $\delta\ev{\hs^x_a} \propto \delta \ev{\hs^z_a \hs^z_b}$ within the triplet state. Similarly, we can trace the calculation of Eq. \eqref{eq:42} as $\delta\ev{\hs^x_a} \leftrightarrow \delta G^{zz;yz}_{-;ab;ab} \leftrightarrow  h^x G^{(0)zy;yz}_{-;ab;ab} \propto \ev{\hs^y_a \hs^y_b}_0$, i.e. $\delta\ev{\hs^x_a} \propto \delta\ev{\hs^y_a \hs^y_b}$. Now we immediately understand that the difference between the two contributions is due to energetics, as it should be. Variation of $\ev{\hs^z_a \hs^z_b}$ has an energy cost $\delta E  \propto J^z \delta \ev{\hs^z_a \hs^z_b}$, whereas for $ \delta\ev{\hs^y_a \hs^y_b}$ there is none at the leading order.

On the other hand, since the algebraic relations are exact, the correct solutions through both Eq. \eqref{eq:39} and \eqref{eq:42} must agree, but that can only be achieved by examining their vertex functions in their SDEOM. By the end of the calculation, we should have a converged expression which accounts for variations in all components of the state. For the above example, the exact induced $\ev{\hs^x_a}$ to the leading order in $h^x$ should receive contributions from all preexisting components of the unperturbed state as
\begin{align}
  \label{eq:45}
  \delta \ev{\hs^x_a} = h_x \left(d_z \ev{\hs^z_a}_0 + d_{yy;ab} \ev{\hs^y_a \hs^y_b}_0 + \dots\right),
\end{align}
where $d_{\alpha\dots}$s are coefficients to be computed dynamically from the unperturbed state.
In general, for a COBS $\{\hu^\alpha\}$, we can write
\begin{align}
  \delta \ev{\hu^{\alpha_0}} = \sum_{\alpha} d_\alpha \ev{\hu^\alpha}_0,
  \label{eq:46}
\end{align}
where $d_\alpha$ should be computed from coefficients of $\delta \hat{H}$ and the unperturbed $G^{(0)}$s. Eq. \eqref{eq:46} can be viewed as the quantum version of variational equations of Hamiltonian systems\cite{inbook-whittaker-1937-treatise} at the wavefunction level.
However, derivation of generic expressions of $d_\alpha$ for an arbitrary $\delta \hat{H}$ shall be discussed elsewhere.

%
%
%
%
%
%
%
%
%
%
%

\section{Algebraic-Dynamical Theory for Lattice Models}
\label{sec:adt-lattice}

For lattice problems, the sizes of $ COBS = \{\hu^\alpha_i, \hT^{\alpha\beta}_{2;ij}, \hT^{\alpha\beta\gamma}_{3;ijk}, \dots \}$
and the corresponding CSDCF grow exponentially with the system size. Therefore, our general goal is not to obtain an exact solution, but rather i) to obtain a quantum statistical description of the ground state with a reasonably small but sufficient COBS and its expectation values, ii) the solutions to the corresponding CSDCF and iii) reliable/robust perturbation theories for finding the ground state as well as the dynamical correlation functions.

In most problems, the local Hamiltonians are composed of short-range two body operators, i.e. grade-2 operators, in some cases grade-3 (such as a chirality term $\chi_{ijk} \propto \s_i \times \s_j \cdot \s_k$ in quantum spin models) or grade-4 (such as a ring-exchange term). Minimization of the energy, i.e. the expectation value of the Hamiltonian, requires minimization (or maximization, depending on the sign of the coupling, but we shall still to minimization regardless) of the grade-2 expectation values. 

These local terms do not always commute. When they do not commute, they cannot be simultaneously minimized. The expectation values of their commutators (the cumulants, irreducible part) put bounds on the minimum. Those commutators would involve operators over longer distance, of higher grade, or both. Such constraints make the expectation values of these new operators relevant for description of the ground state. While how to minimize $\ev{H}$ hence determine the ground state along this line is an important question, we leave discussion of details to future works.

\subsection{Cumulant-Description of States}
\label{sec:cumul-descr-ground}

In this work, we focus on the general dynamical aspects of such static correlations and assume that a sufficiently large set of lattice COBS and the expectation value set can specify a lattice ground state to any desired accuracy in principle:
\begin{align}
  \label{eq:47}
  \ket{GS} \equiv \ket{ \{\ev{\hu^\alpha_i}, \ev{\hT^{\alpha\beta}_{2;ij}}, \ev{\hT^{\alpha\beta\gamma}_{3;ijk}}, \dots \} }.
\end{align}
However, direct usage of the plain $\ev{\hO}$s are inconvenient and mixes many-body correlations and single particle correlation. As found in Sec. \ref{sec:algebr-dynam-theory}, statistically independent coefficients would be more physical and intuitive. For an operator $\hT^{\alpha_1 \alpha_2 \dots \alpha_n}_{n;i_1 i_2 \dots i_n}$ of grade-$n$ with $n>1$,
we make use of the cumulant theory introduced by Kubo\cite{Kubo-1962-gener-cumul} in 1962 and consider its {\it cumulant average} $\ev{\hT^{\alpha_1 \alpha_2 \dots \alpha_n}_{n;i_1 i_2 \dots i_n}}_c$ instead of the plain expectation values. For simplicity, we restrict our discussion up to $n=3$. For example, for bosons we have
\begin{align}
  \label{eq:48}
  \begin{split}
    & \ev{\hu^\alpha_i}_c = \ev{\hu^\alpha_i},\quad \ev{\hu^\alpha_i \hu^\beta_j}_c = \ev{\hu^\alpha_i \hu^\beta_j} - \ev{\hu^\alpha_i} \ev{\hu^\beta_j},\\
    &  \ev{\hu^\alpha_i \hu^\beta_j \hu^\gamma_k}_c = \ev{\hu^\alpha_i \hu^\beta_j \hu^\gamma_k} - \ev{\hu^\alpha_i \hu^\beta_j}_c \ev{\hu^\gamma_k}_c  - \ev{\hu^\alpha_i \hu^\gamma_k}_c\\ &  ~~~\qquad \times \ev{\hu^\beta_j}_c  - \ev{ \hu^\beta_j \hu^\gamma_k}_c \ev{\hu^\alpha_i}_c - \ev{\hu^\alpha_i} \ev{\hu^\beta_j} \ev{\hu^\gamma_k}.
\end{split}
\end{align}
A general expression for the cumulant of an arbitrary $grade-n$ operator is obtained by Meeron\cite{meeron-1957-series-expan}
\begin{align*}
     & \cev{\hT^{\alpha_1  \dots \alpha_n}_{n;i_1 \dots i_n}} = \sum_{l=1}^{n} (-1)^{l-1} (l-1)! \\ & \sum_{\substack{\text{all possible} \\ l~\text{partitions}}} \ev{\hT^{\alpha_{j_1}  \dots }_{n_1;j_1 \dots}}  \ev{\hT^{\alpha_{j_2}  \dots }_{n_2;j_2 \dots}} \dots  \ev{\hT^{\alpha_{j_l}  \dots }_{n_l;j_l \dots}}.
\end{align*}
Here $\{j_1, \dots \} ~ \cup~ \{j_2,\dots\} ~\cup \dots \cup \{j_l,\dots\}$ is a $l$-partition of the sites-set $\{i_1,i_2,\dots,i_n\}$, so are the superscript indices.
$\ket{GS}$ now can be given by a set of statistically independent cumulant averages
\begin{align}
   \label{eq:49}
  \ket{GS} \equiv \ket{ \{\ev{\hu^\alpha_i}, \cev{\hT^{\alpha\beta}_{2;ij}}, \cev{\hT^{\alpha\beta\gamma}_{3;ijk}}, \dots \} }
\end{align}
instead of \eqdisp{eq:43}.

Cumulant averages of elements from a COBS are generally statistically independent unless constrained by constant operator identities. For example, the usual $SU(2)$-spin-$1/2$ COBS $\{\hs^ \alpha \}$ satisfies $(\hs^ \alpha)^2 = 1/4$ and $(\hs^x)^2 + (\hs^y)^2 + (\hs^z)^2 = 3/4$.
However, such information is included or implied already in the closed algebraic relations \eqdisp{eq:3}. For many-body operators of higher grades, such constraints are uncommon due to the exponentially growing degrees of freedom.

\subsubsection{Cumulant correlations as statistical independent degrees of freedom}
\label{sec:cumul-corr-as}
Normally, many-body or composite operators, i.e. operators of grade-$n$ with $n > 1$, only become important with the single particle correlations are gapped out with a substantial gap and a degenerate subspace protected by the gap. A canonical example of such cases is the quantum spins in a Mott insulator. In the Mott phase, all single electron excitations are gapped out by a large gap $U$, i.e. such excitations decay exponentially with a lifetime $\tau \propto 1/U$. Consequently, the electronic spins excitations, which are described by the grade-2 operators $\hS^\alpha = \sum_{ab}\hc^\dagger_a \si_{ab}^\alpha \hc_b$, can be approximately treated as independent degrees of freedom on time scales much larger than $1/U$ or equivalently speaking at small energy scales $\omega \ll U$. These excitations can be viewed as exciting an electron and a hole simultaneously. Should there were no gap, such grade-2 excitations are generally expected to quickly decay into the long-lived single particle excitations.

However, in more complicated strongly interacting systems such as a doped Mott insulator, the situation becomes obscure. Often, we find a plethora of many-body correlations coexist simultaneously which are not necessarily fully gapped out or even gapless.
The analysis of this manuscript shows that, when the ground state possesses such entangled correlations and the SDEOM also have the corresponding dynamical channels, these many-body correlations should be viewed as {\it independent}, both static and dynamical degrees of freedom in general, which are intricately coupled according to their algebraic relations. Such intricate coupling was studied at simplified high-symmetry point with an arguably emergent effective Hamiltonian, such as the $SO(5)$ theory \cite{demler-2004-so-antif-super} or its $SU(4)$ extension\cite{guidry-2020-fermion-dynam}. Within ADT, it is now possible to bridge between microscopic models with such effective theories.

\subsubsection{Cumulants and entanglement}
\label{sec:cumul-entangl}

In general, due to the completeness of COBS, common entanglement measures such as the von Neumann entropy can be expressed in terms of expectation values of COBS in principle. Two well known examples are free fermions and free bosons. Since Wick's theorem is exact, all higher order correlation functions can be factorized, i.e. cumulants for $n >2$ are zero. The reduced density matrices (RDM) of a subsystem $A$ can expressed in terms of $\ev{a_i^{\dagger} a_j}$ for $i,j \in A$, hence the entanglement entropy, etc. can also be expressed in terms of those correlation functions as proved in \refdisp{peschel-2003-calcul-reduc} and its following works.

However, for interacting systems, even weakly interacting systems where the Wick's theorem is presumably still valid, very few analytic results are known\cite{gioev:100503,Ding2011}.
But as pointed out in Sec.~\ref{sec:states}, $\ev{COBS}_c$ can specify an arbitrary density matrix, hence also any RDM, even for strongly interacting systems in principle. Therefore, the ADT approach provides a new way to study the entanglement structures of quantum systems in terms of both static correlations and dynamical properties combined closely.

\subsection{Irreducible Dynamical Correlation Functions and their SDEOM}
\label{sec:irred-dynam-corr}
In accordance to the description of $\ket{GS}$, we introduce the {\it cumulant operators} $\hO^C = \tO$ (we use a $~\tilde{~}~$ to denote the superscript $^C$ for convenience) which satisfying
\ev{\tO} = \cev{\hO}.
For grade-1 operators, apparently $\tu^\alpha_i = \hu^\alpha_i$. For grade-2, we define
\begin{align}
\label{eq:50}
  \tT^{\alpha \beta}_{ij} = (\hu^\alpha_i - \ev{\hu^\alpha_i})(\hu^\beta_j - \ev{\hu^\beta_j}),
\end{align}
which apparently satisfies $\ev{\tT^{\alpha \beta}_{ij}} = \ev{\hu^\alpha_i \hu^\beta_j} - \ev{\hu^\alpha_i} \ev{\hu^\beta_j} = \cev{\hT^{\alpha \beta}_{ij}}$.

Obviously, the cumulant operators also forms a COBS, which we shall call a {\it complete cumulant operator basis} (CCOBS). Note that COBS is independent of the underlying state, but  CCOBS is state-specific.

Now we consider the DCFs of cumulant operators instead, which we shall call {\it irreducible dynamical correlation functions} (iDCFs), denoted as $\tG$, and hence the {\it complete set of iDCFs}(CSiDCF). Correspondingly, the SDEOM of the CSiDCF can cast into a similar matrix form as \eqdisp{eq:16}
\begin{align}
  \label{eq:51}
i \pt_{t_i } [{\mathbf \tG}_{\pm}] - [\mathbf{\tilde{L}}] \cdot [{\mathbf \tG}_{\pm}]= [\mathbf{\tilde{\Delta}}]_{\pm} [\i,\f] .
\end{align}
In this form, we can formally apply the time-domain Fourier transform to solve for $ [{\mathbf \tG}_{\pm}]$ as
\begin{align}
  \label{eq:52}
   [{\mathbf \tG}_{\pm}] = \left( \omega - [\mathbf{\tilde{L}}] \right)^{-1} \cdot [\mathbf{\tilde{\Delta}}]_{\pm}.
\end{align}
\eqdisp{eq:47} has the following implications: i) $[\mathbf{\tilde{L}}]$ encodes the excitation energy levels; ii) $[\mathbf{\tilde{\Delta}}]_{\pm}$ gives the matrix element for each excitation process per the underlying wavefunction or density matrix; iii) \eqdisp{eq:19} is the diagonal form of \eqdisp{eq:48}.

\subsection{Noncanonicality}
\label{sec:noncanonicality}
ADT is a noncanonical theory in nature.
In conventional weak coupling theories, only $G_+$ is considered for bosons and only $G_-$ for fermions. This is due to their canonical nature, which includes two aspects: i) the operators of concern obey canonical commutation or anti-commutation relations, i.e. $[\ho_i,\ho^\dagger_j]_\pm \propto \delta_{ij}$, and ii) Wick's theorem is valid. In strongly interacting systems, either the operators involved only obey noncanonical commutation or anti-commutation relations and automatically invalidates the Wick's theorem, such as the Hubbard operators in $t-J$ models or the quantum spin operators, or the strong correlation itself invalidates the Wick's theorem.
In both situations, such {\it noncanonicality} becomes one of the essential difficulties for any theory.

In ADT, both $G_{\pm}$s are normally necessary for any types of operators for the following reasons.
\begin{enumerate}
\item For completeness, so that the expectation values of any operators, i.e. all components of the state, can be computed from $G_{\pm}$.
\item When deriving effective couplings between emergent degrees of freedom, bosonic type GF ($G_+$) will be needed for fermions and vice versa. For example, in Hubbard model, the superexchange interactions between spins are mediated through the bosonic $G_+$ of the single fermions. This is conventionally done in Lagrangian approaches through the ``integrating out'' procedure. We expect similar procedures in Hamiltonian approaches as well.
\end{enumerate}

The $\delta[\i,\f]_\pm$ in SDEOM of noncanonical $G$s are generally non-local. For example, for bosonic quantum spins, we have $i \pt_t G^{ \alpha \beta }_{-}[\i,\f] \propto \delta(t_i - t_f) \ev{2 \hs_i \hs_f}_c$ while for the canonical GF $G^{ \alpha \beta }_{+}[\i,\f]$, the SDEOM only generate a local term $i \pt_t G^{ \alpha \beta }_{+}[\i,\f] \propto \delta(t_i - t_f) \delta_{i,f} \ev{ \epsilon_{ \alpha \beta \gamma } \hs^\gamma_i} $.

To solve the inhomogeneous SDEOM for the noncanonical GFs, one method is to make use of the traditional ``Green's function method'' by introducing a corresponding $g^{\alpha \beta} _{-}[\i,\f]$ which satisfies
s a homogeneous SDEOM $i \pt_t g^{ \alpha \beta }_{-}[\i,\f] \propto \delta(t_i - t_f) \delta_{\x_i, \x_f}$. Then the original GF can be expressed as its superposition:
  \begin{align}
    \label{eq:53}
    G^{ \alpha \beta }_{-}[\i,\f] = \sum_{j} \ev{2 \hs^\alpha_i \hs^\beta_j}_c g^{\alpha \beta}_{-;j f}[\i,\f] \\
    \rightarrow G^{ \alpha \beta }_{-}[k] = C^{ \alpha \beta }(\k) g^{ \alpha \beta }_{-}[k],\label{eq:54}
  \end{align}
where $C^{ \alpha \beta }(\k)$ is the Fourier transform of $\ev{2 \hs^\alpha_i \hs^\beta_j}_c$. In \eqdisp{eq:50}, we show the expression in $k$-space, which is more convenient.
This understanding shows that ADT can potentially unite and reconcile existing methods and their results such as slave particle, parton, etc.. But we also note that this is not the only way to proceed and can depend on details of the system.

\subsection{Hierarchy of Correlations, Dynamics and Perturbation}
\label{sec:hier-pert}

Anderson raised the concept of {\it emergence} in 1972\cite{anderson-1972-more-is-differ}. The idea was further characterized as a hierarchical structure in correlations from a quantum measurement point of view\cite{aharonov-2018-compl-top}. In ADT, the emergence and the hierarchical structure is reflected in the fact that {\it the cumulant correlations are statistically independent}, possibly up to certain constraints. The higher-grade cumulant average $\cev{\hT^{\alpha_1  \dots \alpha_n}_{n;i_1 \dots i_n}}$ generally {\it cannot} be deduced from the lower grade correlations.

The hierarchical structures of quantum states, or the wavefunctions, have been studied intensively.
A canonical example is the quantum spin correlations in a Mott insulator. As the single electron excitations are fully gapped out, the spin-spin correlations are generally considered independent from the underlying electronic correlation except the strength of superexchange interaction $J$. 
Recently, it is demonstrated by explicit examples\cite{aharonov-2018-compl-top} that higher cumulant correlation functions {\it cannot} be constructed from lower levels such the single particle correlations.

Such hierarchical structures not only exist in quantum states, as reflected by the static or equal time correlations, they are inherited by the dynamical correlations as well.
\eqdisp{eq:51} allows us to examine how the static correlation hierarchy is carried over into dynamics and enable us to understand the perturbation hierarchy demonstrated for the two-spin problem from a more generic perspective. The RHS of \eqdisp{eq:52} is composed of two parts: $(\omega - [\mathbf{\tilde{L}}] )^{-1}$ is essentially the inverse of the excitation spectrum and $[\mathbf{\tilde{\Delta}}]_{\pm}$ reflects the underlying wave-function. The singularities, i.e. the poles, of the correlation functions rely on both.

However, not all correlations are equal. The singularity (poles with weights) must start from the pivotal equations, i.e. what we call the {\it parental channels}. Noting that $\left( \omega - [\mathbf{\tilde{L}}] \right)^{-1}$ is solely determined by the Hamiltonian, such hierarchy is essentially {\it inherited} from the wavefunction or the density matrix. Other correlation functions acquire their singularity through the vertex functions, since their SDEOM provides no direct singularity.


Given the concept of dynamical hierarchy, which is inherited from the correlation hierarchy, the perturbation theory should follow the same hierarchy: perturbative iteration should also begin with the {\it parental channels}.
This is typically not transparent to see, as the subordinate channels are not necessarily ``small" in the usual perturbation sense. In fact, the canonical single particle GFs always have a conserved total spectral weights, hence never appear to be ``small".
The subtlety here is that, as we already discussed in Sec. \ref{sec:analysis-from-adt}, the exact results are unique no matter where the calculation begins. But the routes towards the exact results are diverse. The hierarchical structure provides a tractable way to organize the perturbation series in a better way.


\subsection{Complete Computation Cycles}
\label{sec:pert-comp-scheme}
One of the most important advantages is the capability to formulate mathematically complete cycles of computation, which can have more controllability when approximations are needed.
Given COBS and CSDCF, the state, which is represented as $\ev{COBS}$, can be computed from DCFs, and DCFs can be computed from SDEOM. The completeness guarantees an exact one-to-one mapping  between them.

Consequently, such completeness allows versatile ways to study a strongly interacting system. For example, one can start from a solvable limit, i.e. a exactly solvable ``free'' state, then turn on a perturbation and compute correction to the state and hence the energy. Eventually one can determine the correction to the state by minimizing the energy. One can also start from a trial state, carry out the computation cycles, and require the state and the correlation functions to reach self-consistency.

Most importantly, ADT offers mathematically more rigorous criteria for making controlled approximations which includes i) the entanglement structure and the associated hierarchy and ii) the completeness of components to be kept.
The inherited hierarchical structure of the dynamical correlations and their SDEOM ensures correct perturbation sequences. Such hierarchical structure is essentially the entanglement structure. This understanding is in agreement with many other computational methods, such as density matrix renormalization groups, etc.. The completeness provides a guideline for making truncation to degrees of freedom: the completeness must be preserved in order to have a self-contained calculation.


\section{Discussion}
\label{sec:discussion}
The formulation of the ADT shed new lights on the strongly interacting theories. Below, we discuss the new perspectives put forward by ADT.

The existence of fractionalized excitations is one of the hallmarks of strong correlations. While in rare cases, such as a quasi-holes carrying a fraction of charge $e$ in a Laughlin wave-function, it is possible to construct such fractons explicitly, in most cases, the frationalization is realized through parton or slave particle type of constructions.

However, such constructions are typically biased by priori choices.
For example, for quantum spins, there are Dyson-Maleev boson, Schwinger boson, Holstein-Primakoff boson, Abrikosov pseudo-fermions (also known as Schwinger fermion), etc.. For Hubbard models, there are Kotliar-Ruckenstein slave boson, slave rotor, slave spin
, Hubbard operators representation, etc..
Typically, the Hilbert space is artificially enlarged and the implementation of constraints is difficult.
In addition, relations between physical observables and parton or slave particle correlations are also obscured due to the construction.

Most importantly, different representations can favor one state over the another even for the same problem. For example, for QSMs, bosonic partons are more prone to magnetic orders as Bose-Einstein condensation of the bosons while fermionic partons naturally favors paramagnetic states. Even though it is still possible to describe paramagnetic phases in bosonic theories and vice versa for fermionic theories, additional complexity is often necessary. Moreover, the equivalence or inequivalence between states obtained by different representations is difficult to establish, limiting the relevance of such theories to experiments which require representation-independent results. From the ADT perspective, much of the above mentioned difficulties can be reconciled or understood as we discuss below.

$\bullet$ Different constructions typically are different choices of COBS of the local Hilbert space, which are not necessarily complete and orthogonal.
By considering the complete set of algebraic relations, ADT approach is able to address the many different aspects of strongly interacting systems on equal footing.

$\bullet$ States and physical observables are uniquely determined. The expression of physical observables in terms of functions of auxiliary functions which are analogous to those introduced in parton or slave particle theories but with a more concrete mathematical foundation and are directly related to physical  observables.



\section{Concluding Remarks}
We have formulated a computation framework, the algebraic-dynamical theory, for strongly interacting quantum Hamiltonians on lattices, focusing on systems composed of electrons or/and quantum spins which include Hubbard models, Heisenberg models, Anderson impurity models, periodic Anderson models etc..

By utilizing the complete algebraic relations among a complete operator basis set, we established a complete and consistent relations between a state, characterized by the expectation values of the COBS, and the dynamical correlation functions.
We demonstrated the advantages of ADT through solutions of a two-flavor problem, which can be mapped as the local or mean field limits of the aforementioned models of interest.

For lattice problems, we pointed out that the cumulants of many body correlations essentially encode the entanglement structures and quantum hierarchies. More importantly, the quantum hierarchies are further inherited by the SDEOM of dynamical correlations. Such inherited dynamical hierarchies are likely responsible for many of the difficulties encountered in strong-coupling perturbation theories. Therefore, within ADT, we can use the quantum entanglement structure as a guide for making more controllable approximations. We also discussed the implications of ADT to existing theories.

\section*{Acknowledgements}
We thank R. Yu and J. D. Wu for many beneficial discussions, F. C. Zhang for encouragement and support of this work, and B. S. Shastry for helpful comments on the manuscript. The work at Anhui University was supported by the Anhui Provincial Natural Science Foundation Young Scientist Grant number 1908085QA35 and the Startup Grant number S020118002/002 of Anhui University. WD thanks support from Kavli Institute for Theoretical Sciences for visits during the work.


\bibliography{../../library,../../../org/notes,../../../org/refs}
\end{document}